\documentclass[pra,aps,twocolumn,superscriptaddress,showpacs]{revtex4}
\usepackage{amssymb}

\usepackage{amsmath}
\usepackage{graphicx}
\usepackage{bbm}

\newcommand{\EV}[1]{\langle #1 \rangle}

\newcommand{\rme}{{\rm e}}
\newcommand{\rmi}{{\rm i}}

\newcommand{\ket}[1]{|#1\rangle}
\newcommand{\bk}[0]{{\mathbf k}}
\newcommand{\bq}[0]{{\mathbf q}}
\newcommand{\bn}[0]{{\mathbf n}}
\newcommand{\bm}[0]{{\mathbf m}}
\newcommand{\bx}[0]{{\mathbf x}}

\newcommand{\ddx}[0]{{\rm d}^3x}

\newcommand{\ps}[1]{\hat\psi_{#1}(\bx)}
\newcommand{\psd}[1]{\hat\psi_{#1}^\dagger(\bx)}
\newcommand{\Df}[5]{b_{#1}^\dagger b_{#2} a_{#3,#4}^\dagger a_{#3,#5}}
\newcommand{\Omeff}[0]{\Omega}

\newcommand{\omt}[0]{\omega}
\newcommand{\dt}[0]{\frac{\rm d}{\rm dt}}
\newcommand{\g}[0]{g_{\alpha,\bn,\bn'}^{\bk,\bk'}}

\begin{document}

\title{Fault-Tolerant Dissipative Preparation of Atomic Quantum Registers with Fermions}
\author{A. Griessner}
\affiliation{Institute for Quantum Optics and Quantum Information of the Austrian Academy of Sciences, A-6020
Innsbruck, Austria}
\affiliation{Institute for Theoretical Physics, University of Innsbruck, A-6020
Innsbruck, Austria}
\affiliation{Clarendon Laboratory, University of Oxford, Parks Road, Oxford OX1 3PU,
United Kingdom}
\author{A. J. Daley}
\affiliation{Institute for Quantum Optics and Quantum Information of the Austrian Academy of Sciences, A-6020
Innsbruck, Austria}
\affiliation{Institute for Theoretical Physics, University of Innsbruck, A-6020
Innsbruck, Austria}
\author{D. Jaksch}
\affiliation{Clarendon Laboratory, University of Oxford, Parks Road, Oxford OX1 3PU, United Kingdom}
\author{P. Zoller}
\affiliation{Institute for Quantum Optics and Quantum Information of the Austrian Academy of Sciences, A-6020
Innsbruck, Austria}
\affiliation{Institute for Theoretical Physics, University of Innsbruck, A-6020
Innsbruck, Austria}
\date{25 February 2005}

\begin{abstract}
We propose a fault tolerant loading scheme to produce an array of fermions in an optical lattice of the high
fidelity required for applications in quantum information processing and the modelling of strongly correlated
systems. A cold reservoir of Fermions plays a dual role as a source of atoms to be loaded into the lattice
via a Raman process and as a heat bath for sympathetic cooling of lattice atoms. Atoms are initially
transferred into an excited motional state in each lattice site, and then decay to the motional ground state,
creating particle-hole pairs in the reservoir. Atoms transferred into the ground motional level are no longer
coupled back to the reservoir, and doubly occupied sites in the motional ground state are prevented by Pauli
blocking. This scheme has strong conceptual connections with optical pumping, and can be extended to load
high-fidelity patterns of atoms.
\end{abstract}

 \maketitle

\section{Introduction}

High-precision control of cold atoms in optical lattices has found many potential applications in recent
years, especially in the implementation of quantum information processing and the modelling of strongly
correlated condensed matter systems \cite{hubbardtoolbox}. These applications have been fuelled by
experimental techniques which enable engineering of lattice models with sensitive control over lattice
parameters \cite{Greiner,Esslinger1D,Paredes1D}, independent control for different internal spin states
\cite{sdol}, and control of interactions between atoms via Feshbach resonances
\cite{magfeshbach,optfeshbach}.

For high precision applications, initial state preparation will play a key role in addition to such control
of Hamiltonian parameters \cite{Paeda}. Quantum computing applications generally require an initial register
with exactly one atom per lattice site \cite{Deutschreview}, and observation of interesting effects in
strongly correlated systems often requires the initial spatial patterns of atoms or states with precisely
chosen filling factors \cite{LewensteinKagome}.

The first step in preparation of such states is often adiabatically increasing the lattice potential, making
use of repulsive onsite interactions for bosons \cite{jaksch98} or Pauli blocking for fermions
\cite{Calarcoloading} to load essentially one atom on each lattice site. However, experimental imperfections
will generally lead to non-negligible errors in the resulting states. This can be improved upon by coherently
filtering a state with a filling factor initially greater than one \cite{Paeda,CiracEnsemble}, or potentially
by schemes involving individual addressing and precise measurement of the occupation in individual lattice
sites \cite{BrennenRegister,WeissLoading}. Whilst these methods can, in principle, produce high fidelity
initial states, each of them relies either on the perfect experimental implementation of a single-shot
coherent process or on perfect measurements to avoid defects in the final state. In this article we propose a
\textit{fault-tolerant} loading scheme in which the state being prepared always improves in time. The key
idea is the addition of a dissipative element to the loading process, in contrast to previous schemes, which
rely on coherent transfer or perfect measurements. As we will see below, this dissipative element plays a
similar role in our scheme to that of spontaneous emissions in optical pumping.

Motivated by advances in experiments with cold fermions
\cite{Jin99,Grimm03,Jin03a,Ketterle04,Inguscio02,Esslinger3D,Salomon04,Hulet03,Thomas04}, our scheme is
designed to produce a regular patterned array of fermions in an optical lattice. Fermions have a natural
advantage in initialising atomic qubit registers because Pauli-blocking prevents doubly-occupied sites, and
most of the techniques illustrated using bosons in quantum computing proposals apply equally to fermions.
Fermionic species are also of special interest in the simulation of condensed matter systems
\cite{HofstetterFermiLattice}.

\begin{figure}[htp]
\includegraphics{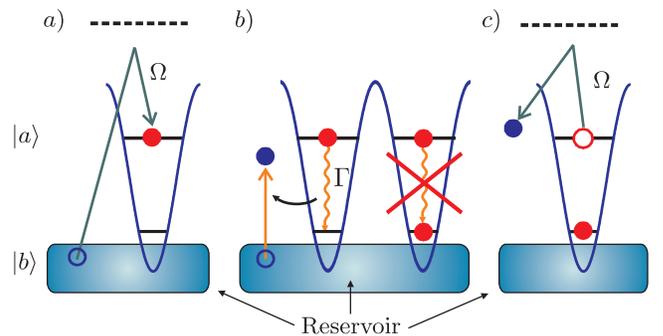}
\caption{Laser-assisted dissipative loading of fermions in an optical lattice: (a) Atoms are coupled from an
external reservoir (in internal state $\ket{b}$) into an excited motional state in the lattice (internal
state $\ket{a}$ via a Raman process; (b) These atoms are cooled to the ground motional level via collisional
interaction with the reservoir atoms, and doubly-occupied sites in the ground level are prevented by Pauli
blocking; (c) Remaining atoms in the excited motional levels are removed by carefully detuning the Raman
coupling above the Fermi Energy. }\label{Fig:setup}
\end{figure}

The setup for our scheme is illustrated in Fig. \ref{Fig:setup}. Atoms in an internal state $\ket{b}$ do not
couple to the lattice lasers, and form a cold Fermi reservoir, which will play the dual role of a source for
atoms to transfer into the lattice, and a bath for cooling lattice atoms. Atoms in the reservoir are coupled
into an excited motional level in the lattice (in internal state $\ket{a}$) via a coherent laser-induced
Raman process (Fig. \ref{Fig:setup}a) \cite{Raizen04}. These atoms are then cooled sympathetically by the
reservoir atoms via collisional interactions, and will decay to the motional ground state together with
creation of a particle-hole pair in the reservoir (Fig. \ref{Fig:setup}b). This is analogous to the
sympathetic cooling process previously presented for a bosonic reservoir in \cite{AJ}. Double occupancy in
the ground state is prevented by Pauli blocking (Fig. \ref{Fig:setup}b), and atoms in the ground state are
not coupled back to the reservoir because the Raman process is far off resonance, so the occupation of the
lowest motional level always increases in time. Additional atoms remaining in excited states at the end of
the process can then be removed by a careful adiabatic detuning and switching off of the coupling lasers
(Fig. \ref{Fig:setup}c).

Such dissipative transfer of atoms into a desired dark state is strongly reminiscent of optical pumping, in
which atoms are excited by a laser, and undergo spontaneous emissions into a desired state which does not
couple to the laser field. The net result of this is to transfer entropy from the atomic system into the
``reservoir'' (the vacuum modes of the radiation field) in order to produce a single pure electronic state
from an initial mixed state. Here the creation of an excitation in the reservoir replaces the spontaneous
emission event, placing the atom in a state where it is not coupled by the Raman process, and leading to the
production of our final pure state, namely a high fidelity array of one atom in each lattice site (or a
pattern of occupied and unoccupied sites).

We note, in addition, that the purely coherent laser-assisted loading could be used as a stand-alone
technique to load the lattice, and could produce high fidelity states if used iteratively, together with
cooling of the Fermi reservoir. Such cooling would fill holes produced in the previous loading step, so that
Pauli blocking would prevent a net transfer of atoms from the lattice to the reservoir, thus ensuring that
the filling factor in the lattice is improved in each step.

The detailed analysis of this dissipative loading process is divided into two parts. Coherent laser-assisted
loading of atoms into the entire lattice in a single addressed motional band is discussed in section
\ref{section:coherent}, and the dissipative transfer of atoms to the lowest motional band is analysed in
section \ref{section:incoherent}. The combination of these two elements into the overall scheme is then
detailed in section \ref{section:combined}.

\section{Laser-Assisted Loading}\label{section:coherent}

We begin by studying the coupling of the atoms forming the reservoir into the optical lattice via a Raman
process, as shown in Fig.~\ref{Fig:setup}. The atoms in the Reservoir are in an internal state $\ket{b}$,
which does not couple to the lasers producing the optical lattice. They form a Fermi gas containing $N$ atoms
with a density $n_{3\rm D}=N/V$ in a volume $V$, with Fermi energy ($\hbar=1$) $\epsilon_F=\left(6\pi^2n_{3
\rm D}\right)^{2/3}/2m$, where $m$ is the mass of the atoms. The internal state $\ket{b}$ is coupled to a
different internal state $\ket{a}$, which is trapped by a deep three dimensional optical lattice potential
$V_a(\bx)$ via a Raman transition.

Our goal is to couple atoms $b$ into the lattice and to achieve an average occupation of fermions close to
one in all lattice sites in one chosen motional band, without coupling to other motional levels. This should
be achieved in a time, where no atoms are allowed to tunnel between different lattice sites and no loss of
atoms occurs (due to, e.g., spontaneous emission events leading to additional internal states etc.).

\subsection{The Model}

The total Hamiltonian of this system is given by
\begin{align}\label{Htot_coh}
    H=H_{a}+H_{b}+H_{\rm RC},
\end{align}
where the Hamiltonians for the atoms $a$ in the optical lattice and for the atoms $b$ forming the reservoir
are
\begin{align}\label{Ha0}
    H_a= \int \ddx \psd{a}\left(-\frac{\nabla^2}{2m}+V_a(\bx)\right) \ps{a},
\end{align}
and
\begin{align}\label{Hb0}
    H_b= \int \ddx \psd{b}\left(-\frac{\nabla^2}{2m}\right) \ps{b},
\end{align}
respectively, in which the anticommuting field operators $\psd{i}$ create a fermion in the internal state
$i\in\{a,b\}$ at the position $\bx$.

\begin{figure}[htp]
\includegraphics{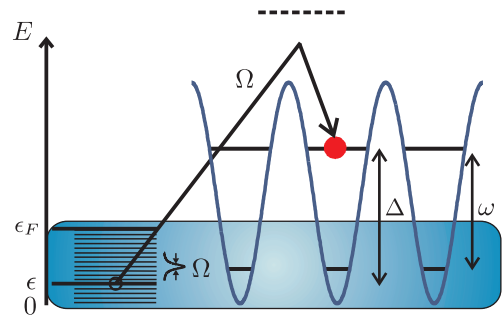}
\caption{Reservoir atoms with energy $\epsilon$ are resonantly coupled to the first excited Bloch band of the
lattice via a Raman laser with two photon Rabi frequency $\Omega$. The resonant energy $\epsilon$ is
experimentally tunable via the Raman detuning, and the energy separation of the two Bloch bands is denoted by
$\omega$.}\label{Fig:setup_2}
\end{figure}

The two internal states are coupled via a Raman process described by the Hamiltonian
\begin{align}\label{Hcoh_fieldops}
    H_{\rm RC}=\int\ddx &\left[\frac{\Omeff}{2}\left( \psd{b}\ps{a} +{\rm h.c.}
    \right) \right.\nonumber \\
    &\left.\quad+\Delta\psd{a}\ps{a}\right],
\end{align}
with the Raman detuning $\Delta$ and the effective (two photon) Rabi frequency $\Omeff$, where we have
assumed running waves with the same wave vectors for two lasers producing the Raman coupling.

We expand the field operators for the free fermions in the reservoir as plane waves and the field operators
for the lattice atoms in terms of Wannier functions,
\begin{align}
    &\ps{b}=\frac{1}{\sqrt{V}}\sum_\bk\rme^{\rmi\bk\bx} b_\bk, \nonumber\\
    &\ps{a}=\sum_{\alpha,\bn}w_\bn(\bx-\bx_\alpha)a_{\alpha,\bn},
\end{align}
where, $b_\bk^\dagger$ creates a reservoir atom with momentum $\bk$, $a_{\alpha,\bn}^\dagger$ is the creation
operator for an atom in lattice site $\alpha$ and motional state with $\bn=(n_x,n_y,n_z)$ in the deep three
dimensional optical lattice, for which $w_\bn(\bx -\bx_\alpha)$, denotes the corresponding Wannier function.

Inserting into Eqs.~(\ref{Ha0})-(\ref{Hcoh_fieldops}) we obtain
\begin{align}\label{Hcoh}
    &H_b=\sum_\bk\epsilon_\bk b_\bk^\dagger b_\bk, \nonumber\\
    &H_a=\sum_{\alpha,\bn} \left(\omega_\bn+\Delta\right) a_{\alpha,\bn}^\dagger
    a_{\alpha,\bn},\nonumber\\
    &H_{\rm RC}=\frac{\Omeff}{2}\sum_{\bk,\alpha,\bn}\left(R_{\bk,\bn}\rme^{-\rmi\bk\bx_\alpha} b_\bk^\dagger
    a_{\alpha,\bn}+ {\rm h.c.}\right),
\end{align}
where the single particle energy of a reservoir atom with momentum $\bk$ is $\epsilon_\bk=|\bk|^2/2m$ and the
energy of a lattice atom in the motional state $\bn$ is given by
\begin{align}
    \omega_\bn=\int\ddx w_\bn(\bx)\left( -\frac{\nabla^2}{2m}+V_a(\bx) \right) w_\bn(\bx).
\end{align}
As we are dealing with very deep optical lattices, tunneling between different lattices sites is strongly
suppressed and has thus been neglected. The Raman coupling parameter $R_{\bk,\bn}$ can be written as
\begin{align}\label{Rkal}
    R_{\bk,\bn}=\frac{1}{\sqrt{V}} \int\ddx \rme^{-\rmi\bk\bx}w_\bn(\bx).
\end{align}

For our deep optical lattices without tunneling between different sites, the periodic lattice is equivalent
to an array of independent microtraps, where each individual trap is well approximated by a harmonic
oscillator. The Wannier functions $w_\bn(\bx-\bx_\alpha)$ can then be approximated by harmonic oscillator
eigenfunctions $\phi_\bn(\bx-\bx_\alpha)$ of the $\bn$-th oscillator level in lattice site $\alpha$. This
approximation allows us to calculate the coupling parameters $R_{\bk,\bn}$ from the Fermi reservoir to the
optical lattice explicitly. For an isotropic three dimensional lattice (where the frequency $\omega$ of each
oscillator is given by $\omega\approx \omega_n-\omega_{n-1}$ and $n\equiv (n_x+n_y+n_z)$) the couplings to
the lowest and first three (degenerate) excited motional states are given by
\begin{align}\label{RkalHO}
    &R_{\bk,0}=\frac{1}{\sqrt{V}}\pi^{3/4} \left(8a_0^3 \right)^{1/2} \rme^{-
    \bk^2 a_0^2/2}, \nonumber\\
    &R_{\bk,1_{x,y,z}}=\sqrt{2} a_0 \rmi \bk_{x,y,z} R_{\bk,0},
\end{align}
where $a_0=\sqrt{1/m\omega}$ denotes the size of the harmonic oscillator ground state, and the index $x,y,z$
labels the coupling to the three degenerate states of the first excited oscillator level.

The characteristics of the coherent loading procedure strongly depend on the interplay between the
(experimentally adjustable) parameters: the detuning $\Delta$ and two photon Rabi frequency $\Omeff$ of the
lasers producing the Raman coupling, the Fermi energy $\epsilon_F$ and the separation $\omega$ of the
oscillator levels. The Raman detuning can be adjusted to address different states in the Fermi sea and
different motional states in the lattice. In the following we write $\Delta = -5\omega/2 + \epsilon$ to
indicate the resonant coupling of reservoir atoms with energy $\epsilon$ to the $n=1$ motional states of each
lattice site. We note that it is straightforward to address other motional states in the lattice (e.g. to
directly load the lowest level) by adjusting the detuning $\Delta$. However, as we will later use the
transfer of the atoms from the reservoir to the lattice as a first step of an indirect loading of the lowest
motional states as described in the introduction, we choose the transfer to the first excited motional state
here. To be able to selectively fill the first excited oscillator levels, the conditions
$(\epsilon_F-\epsilon) \ll \omega$ and $\epsilon \ll \omega$ (and consequently $\epsilon_F\ll\omega$) have to
be fulfilled, in order to avoid unwanted coupling to higher excited and to the lowest motional state,
respectively.

\subsection{The Fast and Slow Loading Regimes}

The physics of the loading process allows us to identify two different loading limits: (1) the ``fast loading
regime'', where
\begin{align}\label{fastOmegacond}
    \Omeff\gg \omega,\epsilon_F,
\end{align}
and (2) the ``slow regime'', where
\begin{align}\label{slowOmegacond}
    \Omeff\ll \omega,\epsilon_F.
\end{align}
Below we will see that our goal to selectively fill a certain motional state without coupling to other states
can only be achieved in the slow loading regime, but to obtain more insight into the physics of the loading
dynamics it is instructive to discuss both regimes.

In the fast loading regime, where the Rabi frequency $\Omeff$ is the largest frequency scale in the system,
the loading is performed in a very short time $T\sim \pi/\Omeff \ll a_0/v_F$, with $v_F=
\sqrt{2\epsilon_F/M}$ the Fermi velocity, where atoms in the Fermi reservoir do not move significantly during
the loading on a lengthscale given by the size $a_0$ of the harmonic oscillator ground state. The Wannier
modes in the lattice then couple to localized reservoir fermions at each site, and thus the dynamics for
different sites decouple. Given there is at least one fermion in the reservoir per size $a_0$ of the ground
state in each lattice site during the loading, i.e., given the density of the reservoir atoms
\begin{align}\label{denscondfast}
    n_{3  \rm D}\gtrsim 1/a_0^3,
\end{align}
each $n=1$ motional state in each lattice site can be filled with at least one atom from the reservoir by
applying a $\pi$-pulse $\Omeff T=\pi$. For an optical lattice with $\omega/2\pi\sim 50$kHz the required
densities of the Fermi gas are $n_{3\rm D}\gtrsim 3\times 10^{15}$cm$^{-3}$ for $^{40}$K and for deeper
lattices the required densities are even higher. The condition (\ref{denscondfast}) for the density of the
reservoir can be expressed in terms of energies as
\begin{align}\label{fastepsfom_condition}
    \epsilon_F \geq (6\pi^2/\sqrt{2})^{2/3} \omega.
\end{align}
This inequality violates the condition $\epsilon_F\ll\omega$, which is necessary to be able to selectively
address individual motional states. Consequently, unwanted population will be transferred to additional
motional states in this loading limit, which would have to be carefully removed after the loading process.

In the slow loading regime, where condition (\ref{slowOmegacond}) is fulfilled, the atoms in the Fermi
reservoir are no longer frozen during the loading process, but are allowed to move with respect to the
lattice during the loading. This is now performed in a time $T\gg \lambda/2v_F$, where $\lambda/2$ is the
lattice spacing. Consequently, the density condition (\ref{denscondfast}) can be relaxed to
\begin{align}\label{denscondslow}
    n_{3\rm D}\left(\frac{\lambda}{2}\right)^3\gtrsim 1,
\end{align}
i.e., we only need one atom in the reservoir per lattice site to be able to efficiently fill the lattice. For
typical experimental parameters $\lambda\sim 800$nm for $^{40}$K this results in the condition $n_{3\rm
D}\gtrsim 10^{13}$cm$^{-3}$, which has already been achieved in current experiments (e.g.~\cite{Jin03}). The
density condition (\ref{denscondslow}) expressed in terms of energies now reads
\begin{align}\label{slowepsFomRcond}
    \epsilon_F\gtrsim \left(\frac{6}{\pi}\right)^{2/3} \omega_R,
\end{align}
with $\omega_R=2\pi^2/m\lambda^2$ the recoil frequency. As $\omega_R\ll\omega$ for a deep optical lattice,
the condition $\epsilon_F\ll\omega$ can be fulfilled in this loading limit, and as $\Omeff\ll\omega$,
individual motional states in each site can be addressed. In the following we will investigate these two
extreme limits and the intermediate regime in detail.

\subsection{Analysis of the Loading Regimes}

\subsubsection{Fast Loading Regime}

In this regime, where the motion of the atoms in the reservoir is frozen on the scale $a_0$ during the
transfer, the physics is essentially an on-site coupling and transfer. We thus find it useful to expand the
modes in the reservoir in terms of localized Wannier functions corresponding to the lattice. Such an
expansion of the reservoir modes arises naturally from the definition of the matrix elements $R_{\bk,\bn}
\rme^{-\rmi\bk\bx_\alpha}$ (Eq.~(\ref{Rkal})) and allows us to write
\begin{align}\label{HRCwannier}
    H_{\rm RC}=\frac{\Omeff}{2}\sum_{\alpha,\bn}\left( B_{\alpha,\bn}^\dagger a_{\alpha,\bn}+{\rm h.c.}\right),
\end{align}
where $B_{\alpha,\bn}=\sum_\bk R_{\bk,\bn}b_\bk \rme^{-\rmi\bk\bx_\alpha}$ is the mode corresponding to the
Wannier function $w_\bn(\bx-\bx_\alpha)$. Note that these collective modes fulfill
\begin{align}\label{Borth}
    \{B_{\alpha,\bn},B_{\beta,{\bf m}}^\dagger\}=\delta_{\alpha,\beta}\delta_{{\bf m},\bn},
\end{align}
(where $\delta$ denotes the Kronecker Delta), i.e., modes corresponding to different lattice sites or to
different motional states are orthogonal. Furthermore, in the fast regime we can neglect the first two terms
$H_a$ and $H_b$ in the Hamiltonian (\ref{Htot_coh}) due to the condition Eq.~(\ref{fastOmegacond}) during the
loading time $T\sim\pi/\Omeff$ and the total Hamiltonian can be approximated by $H \approx H_{\rm RC}$. The
sites thus decouple, and the loading process at each site proceeds independently, but with the same Rabi
frequency $\Omeff$ for the coupling.

We are interested in the time evolution of the matrix elements of the single particle density matrix, i.e.,
$\EV{a_{\alpha,\bn}^\dagger a_{\beta,\bm}}$, $\EV{a_{\alpha,\bn}^\dagger B_{\beta,\bm}}$ and
$\EV{B_{\alpha,\bn}^\dagger B_{\beta,\bm}}$. In the fast loading regime, where $H\approx H_{\rm RC}$, the
respective matrix elements can be calculated analytically from the Schr\"odinger equation with the
Hamiltonian Eq.~(\ref{HRCwannier}), and we find for states, where $\Omeff/2 \gg |(n-1)\omega|$, i.e.,
$t\lesssim T\ll a_0/v_F$
\begin{align}\label{TEoccfast}
    &\EV{a_{\alpha,\bn}^\dagger a_{\beta,\bm}(t)}=\delta_{\alpha,\beta}\delta_{\bn,\bm} \sin^2\frac{\Omeff}{2}t,
\end{align}
and
\begin{align}
    &\EV{B_{\alpha,\bn}^\dagger B_{\beta,\bm}(t)}=\delta_{\alpha,\beta}\delta_{\bn,\bm} \cos^2\frac{\Omeff}{2}t,
\end{align}
for the time evolution of the occupation of the modes in the lattice and in the Fermi sea, respectively.
These expressions assume that the lattice modes are initially empty and the corresponding modes in the Fermi
sea are initially filled. If the Fermi sea is initially filled up to $\epsilon_F$, then this assumption is
fulfilled for any $\alpha$ and $\bn$ for which that each mode $B_{\alpha,\bn}$ contains contributions only
from states with energy below $\epsilon_F$. Thus, in the fast loading regime the occupation in the lowest and
first excited motional state undergoes Rabi-oscillations at a Rabi frequency $\Omeff$, and provided the
density is sufficiently high, the lattice can be efficiently filled by applying a $\pi$-pulse,
\begin{align}\label{pipulse}
    \Omeff T \sim\pi,
\end{align}
with loading time $T$. Atoms will also be coupled to other motional states in the lattice with the resulting
filling factors depending on the density of the reservoir gas and the actual value of the Rabi frequency
$\Omeff$.

\begin{figure}[htp]
\includegraphics{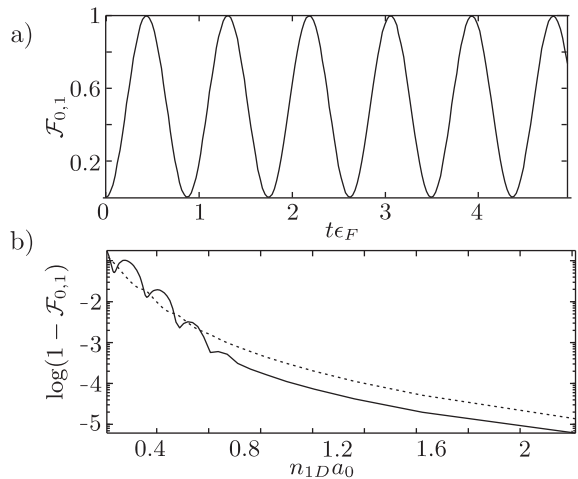}
\caption{Numerical results in the fast loading limit. (a) The time evolution of the occupation of the lowest
($n=0$) and first excited motional ($n=1$) levels against time in dimensionless units. Note that these lines
are indistinguishable. (b) $1-\mathcal{F}_n(t=\pi/\Omega)$ after applying a $\pi$-pulse, for the lowest
($n=0$, dotted line) and the first excited ($n=1$, solid line) motional state, as a function of the
dimensionless density $n_{\rm 1D}a_0$. Parameters used: N=201 particles in the Fermi sea, M=5 lattice sites,
$\Omeff=17.8\epsilon_F$, $\Delta=-3\omega/2$ and in (a) $\omega=0.1~\epsilon_F$, whereas in (b) $\omega$ is
varied.}\label{Fig:fast}
\end{figure}

To model the full loading dynamics we use numerical simulations of the dynamics generated by the Hamiltonian
(\ref{Htot_coh}). In these simulations we only consider the lowest two motional states for simplicity, but
all results are easily extended to more motional states. Also, the simulations are one dimensional, which
means that the excited oscillator state with $n=1$ is no longer degenerate. Because couplings to motional
excitations in different spatial directions are independent, such simulations are representative for loading
into each of the three 3D modes.

In Fig.~\ref{Fig:fast}a we show the results of our numerical simulations of the complete system described by
the Hamiltonian Eq.~(\ref{Htot_coh}) in the fast loading limit. In the upper and lower part we plot the
fidelity $\mathcal{F}_m(t)\equiv\sum_\alpha \EV{a_{\alpha,m}^\dagger a_{\alpha,m}(t)}/M$, with $M$ the number
of lattice sites, of the lowest and first excited Bloch band as a function of time in dimensionless units
$t\epsilon_F$. The numerical results are in excellent agreement with the analytical calculations
(Eq.~(\ref{TEoccfast})), as we find oscillations of the fidelity in both Bloch bands between zero and
$\mathcal{F}_m(t) \gtrsim 1-10^{-4}$ occur with a Rabi frequency $\Omeff$. In Fig.~\ref{Fig:fast}b we analyze
the scaling of the fidelity in the two bands with the dimensionless density $n_{3\rm D} a_0^3$ (i.e., with
$n_{1D} a_0=\sqrt{2\epsilon_F/\omega}/\pi$ in our one dimensional simulations, with $n_{1D}$ the one
dimensional density of the reservoir gas). As expected, the fidelity after a $\pi$ pulse, i.e.,
$\mathcal{F}_m(t=\pi/\Omeff)$ increases with the density, and high fidelity states can be achieved for large
densities $n_{1D} a_0\gtrsim 1$. In this and all numerical simulations below we have checked that the results
are independent of the quantization volume, which is much smaller than in a real experiment, due to the
comparably small number of particles in the simulations.

\begin{figure}[htp]
\includegraphics{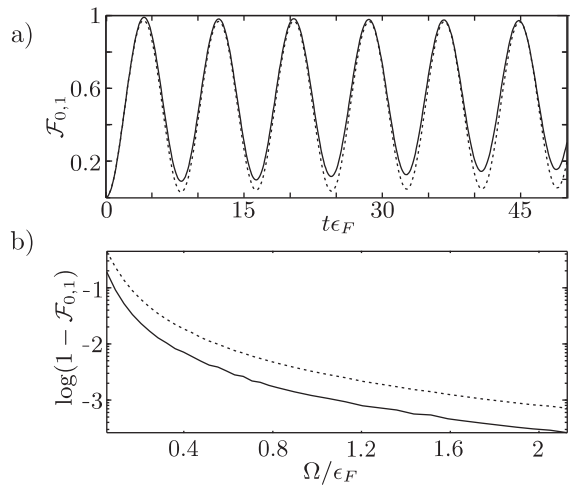}
\caption{Numerical results for $\Omega$ approaching an intermediate regime from the fast loading limit. (a)
The time evolution of the occupation of the lowest ($n=0$, dotted line) and first excited ($n=1$, solid line)
motional state against time in dimensionless units, again for $N=101$, $M=5$ and for a typical set of
parameters $\Omeff=0.72 \epsilon_F$, $\Delta=-3\omega/2$ and $n_{\rm 1D}a_0\sim 2$. (b) The occupation number
for the lowest ($n=0$, dotted line) and first excited ($n=1$, solid line) after applying a $\pi$-pulse versus
the dimensionless Rabi frequency $\Omeff$. }\label{Fig:fastOmegascaling}
\end{figure}

In Fig.~\ref{Fig:fastOmegascaling} we show how the loading dynamics change when approaching the intermediate
regime from the fast limit, i.e. the scaling of the fidelity with the Rabi frequency $\Omeff$. In
Fig.~\ref{Fig:fastOmegascaling}a we show the qualitative behaviour of the loading dynamics for typical
parameters, in Fig.~\ref{Fig:fastOmegascaling}b the scaling of the fidelity $\mathcal{F}_m (t=\pi/\Omega)$,
$m=0,1$ is shown as a function of the Rabi frequency. These numerical simulations show that the Bloch bands
still cannot be individually addressed, and the fidelity becomes worse if the Rabi frequency is decreased.

Thus, our chosen motional state can, in principle, be efficiently filled in this regime on sufficiently fast
timescales. However, the requirements on the density are difficult to achieve experimentally, and occupation
in other motional states cannot be avoided. As a result in this regime we obtain no significant advantage
over traditional loading mechanisms such as adiabatically turning on the lattice. In the next section we will
investigate the slow loading regime. In this limit these problems do not exist and we are able to selectively
load a single energy level efficiently.

\subsubsection{Slow Loading Regime}

In this regime, transport is significant during the loading, and the system dynamics are described by the
complete Hamiltonian~(\ref{Htot_coh}). As the reservoir atoms move between Wannier modes during the loading
process, it is now more convenient to directly use the momentum representation Eq.~(\ref{Hcoh}) to express
the coupling Hamiltonian.

From Eq.~(\ref{Hcoh}) one can see that each lattice site $\alpha$ and each motional state $\bn$ is coupled to
many momentum modes $b_\bk$ in the reservoir. However, as $\Omeff\ll\epsilon_F$, effectively only a subset of
momentum modes with energies centered around the resonant frequency $\epsilon = \Delta+3\omega/2$ is coupled
to the lattice, whereas the remaining states are far detuned and the transfer is suppressed. The width of
this effective coupling range depends on both the Rabi frequency $\Omega$ and the matrix elements
$R_{\bk,\bn}$, and an upper bound for the width of this range is given by the Rabi frequency $\Omega$.

It is convenient to rewrite the coupling Hamiltonian of Eq.~(\ref{Hcoh}) as
\begin{align}\label{HRCslow}
    H_{\rm RC}=\sum_{\bk,\bn} \left[ R_{\bk,\bn} b_\bk^\dagger \left(\sum_\alpha \rme^{-\rmi\bk\bx_\alpha}
    a_{\alpha,\bn}\right) +{\rm h.c.}\right],
\end{align}
from which we can see that each momentum mode in the reservoir couples to a collective mode $\sum_\alpha
\rme^{\rmi \varphi_{\bk,\alpha}}a_{\alpha,\bn}$ in the lattice. To fill the lattice it is necessary that the
range of states in the reservoir couples to at least $M$ orthogonal collective modes in the lattice. Writing
the phase as
\begin{align}
    \bk\bx_\alpha=\pi\sqrt{\frac{\epsilon_F}{\omega_R}}\left(\frac{\bk}{k_F} \frac{\bx_\alpha}{\lambda/2}
    \right),
\end{align}
we see that it is necessary to couple a range of states with width of at least $k_F\sqrt{\omega_R/
\epsilon_F}$ in momentum space to the lattice to fill $M$ lattice sites. In the slow regime, where $\Omega\ll
\epsilon_F$ and furthermore $\omega_R\lesssim \epsilon_F$ (from the density condition
(\ref{slowepsFomRcond}), the recoil frequency will typically exceed the Rabi frequency, i.e., $\Omega <
\omega_R$. As only states within a range $\epsilon \pm \Omega$ are coupled to the lattice, the lattice cannot
be filled efficiently for a constant $\epsilon$.

Thus to achieve a high population in the desired motional state of each lattice site we must sweep the
resonant frequency $\epsilon$ through a range of at least $\omega_R$, scanning through many modes. Such a
procedure also has the advantage that as we only couple to a narrow range in the Fermi sea at any one time,
the reverse process of transferring particles from the lattice to the Fermi sea will be suppressed by Pauli
blocking. In our numerical simulations we linearly sweep the detuning from $\epsilon=0$ to
$\epsilon=\epsilon_F$ in a loading time $T$.

We are interested in the time evolution of the matrix elements  $\EV{a_{\alpha,n}^\dagger a_{\beta,m}}$,
$\EV{a_{\alpha,n}^\dagger b_\bk}$, and $\EV{b_\bk^\dagger b_{\bk'}}$ of the single particle density matrix.
For a system described by a quadratic Hamiltonian the equations of motion for the second order correlation
functions can be obtained from the (linear) Heisenberg equations (see Appendix \ref{Heisenberg}). As the
system is described by the quadratic Hamiltonian (\ref{Htot_coh}) and (\ref{Hcoh}), the linear Heisenberg
equations for the operators $a_{\alpha,n}$ and $b_\bk$ have the simple form (again only considering the
lowest two motional states in a one dimensional system)
\begin{align}
    &\dot a_{\alpha,1}= -\rmi \frac{\Omeff}{2}\sum_\bq R_{\bq,1}^* \rme^{\rmi\bk\bx_\alpha} b_\bq - \rmi
    \epsilon a_{\alpha,1},\nonumber\\
    &\dot b_\bk= -\rmi\frac{\Omeff}{2}\sum_{\mu,n} R_{\bk, n} \rme^{-\rmi\bk\bx_\mu} a_{\mu,n} -\rmi
    \epsilon_\bk b_\bk, \nonumber\\
    &\dot a_{\alpha, 0}=-\rmi \frac{\Omeff}{2}\sum_\bq R_{\bq,0}^* \rme^{\rmi\bk\bx_\mu} b_\bq+
    \rmi(\omega-\epsilon) a_{\alpha,0},
\end{align}
which can be used to efficiently calculate the time evolution of the desired functions numerically. Note that
in an isotropic three dimensional lattice again all three degenerate $n=1$ states will be loaded by sweeping
the resonant frequency $\epsilon$ through the Fermi sea. In practice it is also possible to selectively load
only a single atom in each lattice site by shifting two excited motional states out of resonance, choosing an
anisotropic lattice with significantly higher oscillator frequencies in two dimensions.

\begin{figure}[htp]
\includegraphics{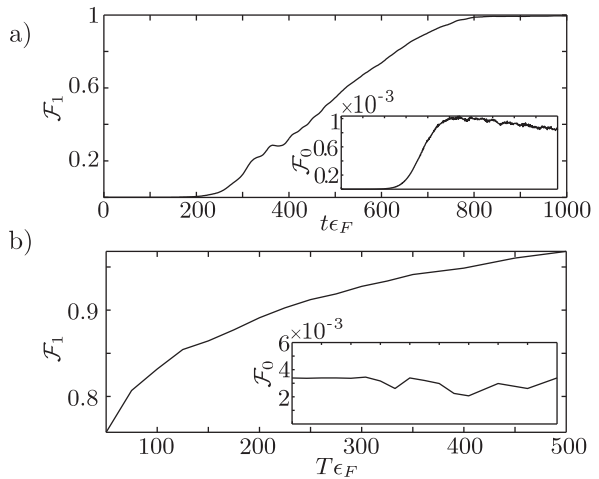}
\caption{Numerical simulation of the loading dynamics in the ``slow loading'' limit. (a) Occupation of the
lowest and first excited motional bands as a function of time, showing attainment of a high fidelity in the
excited band. The resonant frequency $\epsilon$ is swept from $\epsilon_F \rightarrow 0$, and $\Omega$ is
ramped from $0$ to $0.45 \epsilon_F$, reaching that value at $t\epsilon_F=500$. Parameters used: $N=81$
particles in the Fermi sea, $M=5$ lattice sites, $\omega=5\epsilon_F$ and $n_{\rm 1D}\lambda/2=3.4$. (b) The
final occupation number after a loading sweep with constant $\Omega$, and $\epsilon$ ramped from $\epsilon_F
\rightarrow 0$, versus the dimensionless sweep time $\epsilon_F T$. Parameters used: $N=81$, $M=5$,
$\Omega=0.9 \epsilon_F$ and $\omega=10\epsilon_F$ and $n_{\rm 1D}\lambda/2=1.7$.}\label{Fig:slowgeneric}
\end{figure}

In Fig.~\ref{Fig:slowgeneric}a we show numerical results for the time evolution of the occupation number in
the first (upper plot) and in the lowest (lower plot) Bloch band as a function of time in dimensionless
units. Here, $\Omega$ is slowly switched on to reduce the additional holes introduced in the Fermi sea by
coupling atoms into states above $\epsilon_F$. This is an example of many possible optimisations to produce
high filling, and we find the final $\mathcal{F}_1> 0.99$, in a time of the order of $10$ milliseconds (with
$\omega \sim 2\pi \times 100$ kHz). In Fig.~\ref{Fig:slowgeneric}b the occupation of the two motional levels
after a loading sweep is plotted as a function of the sweep time $T$. These results are not optimised
($\Omega$ is held constant, and we sweep $\epsilon$ from $\epsilon_F \rightarrow 0$), but still produce
fidelities $\mathcal{F}_1>0.95$ on a timescale of a few milliseconds, and we see that the average filling
factor increases with the loading time.

It is important to note that whilst high fidelities can be obtained by optimising the parameters of the
sweep, it is not necessary to achieve high filling during this sweep in order to produce high fidelities for
the overall loading scheme. In the full scheme with decay of atoms to the ground motional state included, the
upper band need never be completely filled at any one time, and removal of atoms via the decay process will
lead to further atoms being coupled into the lattice in the upper motional band.

\begin{figure}[htp]
\includegraphics{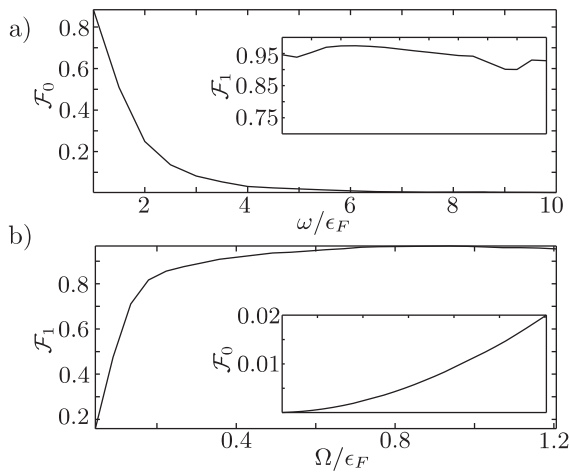}
\caption{In (a) we show the occupation of the lowest and first excited Bloch band after a linear loading
sweep from $\epsilon=\epsilon_F\rightarrow 0$ with $\epsilon_FT=300$ versus the band separation $\omt$ in
dimensionless units. Part (b) shows the loading dynamics approaching an intermediate regime from the slow
regime: We plot the occupation of the lowest and first excited Bloch band after a linear loading sweep, with
$\epsilon_FT=300$ against $\Omega$. Parameters used: N=81 particles in the Fermi Sea, M=5 lattice sites and
$n_{\rm 1D}\lambda/2=1.7$, in (a) $\Omega=0.9\epsilon_F$ and in (b)
$\omt=10\epsilon_F$.}\label{Fig:slowscaling}
\end{figure}

Due to the condition $\Omeff \ll \omega$, unwanted coupling to other Bloch bands can be avoided in this
regime, by choosing $\epsilon \ll \omega$ (c.f. Fig.~\ref{Fig:setup}), as the coupling is then sufficiently
far detuned as demonstrated in the lower two plots of Fig.~\ref{Fig:slowgeneric}. The scaling of the unwanted
coupling to the lower band is shown in Fig.~\ref{Fig:slowscaling}a, where we plot the occupation of the two
Bloch bands after a linear sweep with $\epsilon_F T=300$ against the ratio $\omega/\epsilon_F$.

In Fig.~\ref{Fig:slowscaling}b we show the numerical results when approaching the intermediate regime, i.e,
the scaling of the occupation of the two bands after the linear sweep with the Rabi frequency. We find that
also here high occupation of the first Bloch band can be achieved, but by increasing the Rabi frequency the
unwanted coupling to the lower band also increases, as can be seen in the lower plot of the figure.

In summary, in the ``slow loading'' regime high fidelity loading of the $n=1$ motional level can be achieved
on timescales much shorter than those on which atoms are lost from the lattice by sweeping the resonant
coupling frequency $\epsilon$ through the Fermi sea. This loading mechanism gives us the significant
advantage over simple loading techniques such as adiabatically increasing the lattice depth that we can
address a particular energy level in the lattice, whilst not coupling to levels that are sufficiently far
detuned. This property can also be used to load patterns of atoms, because if a superlattice is applied, then
the energy of certain lattice sites can be shifted out of resonance with the Raman process, so that no atoms
are coupled into these sites.

In the next section we will discuss the cooling of atoms in higher motional levels to the ground state, which
removes the atoms from the motional state being coupled from the reservoir. Together with Pauli blocking of
modes in the lattice, this allows us to make the overall loading process fault-tolerant. As an additional
remark, though, we note that this laser-assisted loading of a selected energy level in the lattice could be
used as a stand-alone technique to load the lattice, e.g., coupling atoms directly into the ground motional
state. (In order to load an excited motional state in this manner, interaction of atoms in the lattice and
atoms in the reservoir must be made very small on the timescale of the loading process, e.g., by using a
Feshbach resonance, in order to avoid decay of the atoms into the ground state). This process on its own is
not as robust as the procedure we obtain by including a dissipative element in the loading scheme, which will
be discussed in the next section. However, reasonably high fidelities could still be obtained with this
method alone, especially if the method was applied iteratively, cooling the Fermi reservoir between each two
steps. Net transfer of atoms already in the lattice back to the reservoir would be prevented in each step by
Pauli blocking in the filled Fermi sea. Note again that as with the full dissipative loading scheme, a single
sweep would also not need to completely fill the upper band. The dissipative element discussed in the next
section allows for the production of an arbitrarily high-fidelity state without the requirement of
iteratively cooling the Fermi reservoir.

\section{Dissipative Transfer: Cooling Atoms to the Lowest Band}\label{section:incoherent}

The second stage of the loading process is cooling atoms in an excited motional state to the ground state via
interaction with the reservoir gas. This is closely related to the cooling process with a bosonic reservoir
in \cite{AJ}. The external gas here plays the role of an effective $T=0$ heat bath for the lattice atoms, and
ground state cooling is achieved on timescales much shorter than atoms are lost from the lattice.

We consider the coupling of lattice atoms $a$ via a collisional interaction to the atoms $b$ in the reservoir
so that the system is described by the Hamiltonian
\begin{align}\label{Htotinc}
    H=H_a+H_b+H_{\rm int},
\end{align}
where the collisional interaction, $H_{\rm int}$, between two fermions is the usual density-density
interaction
\begin{align}\label{Hinc0}
    H_{\rm int}=g\int\ddx\psd{a}\ps{a}\psd{b}\ps{b},
\end{align}
with $g=4\pi a_s/m$ and $a_s$ the $s$-wave scattering length. Expanding the field operators as described in
the previous section we obtain
\begin{align}\label{Hint}
    H_{\rm int}=\sum_{\bk,\bk'\atop\alpha,\bn,\bn'} \g \Df{\bk}{\bk'}{\alpha}{\bn}{\bn'},
\end{align}
which is local in each lattice site because of the small overlap between Wannier functions for neighboring
sites in a deep lattice, with
\begin{align}
    \g= \frac{g}{V}~\rme^{\rmi \bx_\alpha(\bk'-\bk)} \int\ddx \rme^{\rmi\bx(\bk'-\bk)}
    w_\bn(\bx) w_{\bn'}(\bx).
\end{align}
Each $\g$ describes a scattering process in which a particle-hole pair is created in the reservoir by
scattering an atom from momentum state $\bk\rightarrow\bk'$, combined with the transition of an atom at site
$\alpha$ from motional state $\bn\rightarrow\bn'$.

If the transition in the lattice is from a higher energy mode to a lower energy mode, this corresponds to a
cooling transition, whereas the reverse process constitutes heating. As the initial temperature of the
reservoir $k_B T\ll \epsilon_F\ll\omega$, the heating processes will be, at least initially, insignificant,
as few reservoir atoms will exist with sufficient energy to excite an atom in the lattice. If the number of
atoms in the reservoir is large compared to the number of sites in the lattice ($N\gg M$), then the rate of
heating processes due to interaction with previously excited atoms will be small compared to cooling
processes due to interaction with atoms remaining below the Fermi energy $\epsilon_F$. Because the cooling
processes in different lattice sites couple to different modes, and therefore are incoherent, the reservoir
can then be treated throughout the process approximately as a $T=0$ bath.

This can be further enhanced in two ways. Firstly, in an experiment in which the reservoir gas is confined in
a weak harmonic trap, particles with sufficiently large energies can be allowed to escape from the trap. The
large separation of the Bloch band $\omega$, and corresponding excitation energy will then cause many excited
reservoir atoms to leave the trap, providing effective evaporative cooling during the process. Secondly, the
lattice depth could be modulated during the experiment, so that the excitation energy changes, decreasing the
probability that atoms are heated by previously excited reservoir atoms.

The cooling dynamics are then described in the Born-Markov approximation by a Master equation for the reduced
density operator $\rho$ for the atoms in the lattice. If we consider coupling of atoms from the first excited
motional levels $\bn\in \{(1,0,0),(0,1,0),(0,0,1)\}$ to the ground state, the resulting master equation
(derived in Appendix \ref{ME}) is
\begin{align}\label{MEalbet}
    \dot\rho=\sum_{\alpha,\beta,\bn}\frac{\Gamma_{\alpha,\beta,\bn}}{2} \left(2A_{\alpha,\bn} \rho
    A_{\beta,\bn}^\dagger -A_{\alpha,\bn}^\dagger A_{\beta,\bn} \rho- \rho A_{\alpha,\bn}^\dagger
    A_{\beta,\bn}  \right),
\end{align}
with
\begin{align}\label{Gamma}
    \Gamma_{\alpha,\beta,\bn}=&2\pi \sum_{\bk, \bk' \atop k'>k} g_{\alpha,\bf{1},\bf{0}}^{\bk,\bk'}
    {g_{\beta,\bf{1},\bf{0}}^{\bk,\bk'}}^* \delta \left(\omega-\epsilon_\bk+ \epsilon_{\bk'} \right) \nonumber\\
    &\approx \frac{g^2n_{3 \rm D}m}{\pi a_0\sqrt{2}}\frac{2}{3e}~\delta_{\alpha, \beta}.
\end{align}
Here, the jump operator $A_{\alpha,\bn}= a_{\alpha,0}^\dagger a_{\alpha,\bn}$ describes the cooling of a
lattice atom in site $\alpha$ from the first excited motional level $\bn$ to the ground state. These results
are obtained by calculating the integral over momenta in the Fermi sea to lowest order in
$\epsilon_F/\omega$.

The approximation in the second line of Eq.~(\ref{Gamma}), in which neglect off diagonal terms $\alpha
\not=\beta$ amounts to the approximation that the coherence length of the Fermi reservoir is much shorter
than the lattice spacing. This is true provided that the wavelength of the emitted particle excitation,
$\sqrt{2\pi^2/(m\omega)}$, is much shorter than the lattice spacing, i.e., $\omega_R/\omega \ll 1$. This is
consistent with the previous approximation that the lattice is so deep that we can neglect tunnelling between
neighbouring sites. This can be seen directly when these off-diagonal terms are calculated, as for large
$\omega_R/\omega$ they decay (to lowest order in $\epsilon_F\ll\omega$) as
\begin{equation}
\Gamma_{\alpha,\beta,\mathbf{n}}\sim\frac{\sin (\pi\sqrt{\omega/\omega_R}
|\alpha-\beta|)}{\pi\sqrt{\omega/\omega_R} |\alpha-\beta|}.\label{sincdecay}
\end{equation}
This effect is analogous to the spontaneous emission of two excited atoms which are separated spatially by
more than one wavelength of the photons they emit. In this case, the atoms can be treated as coupling to two
independent reservoirs, and effects of super- and sub-radiance do not play a role.

For typical experimental values $n_{3\rm D}\sim 10^{14}{\rm cm}^{-3}$ and $a_s=174a_B$, for $^{40}$K as given
in \cite{Jin03}, with the Bohr radius $a_B$ and a deep optical lattice with $\omega/2\pi \sim 100$kHz, we
find a decay rate $\Gamma /2\pi \sim 3.6$kHz. Thus, cooling can again be achieved fast enough, as this rate
is much faster than typical loss rates of the lattice atoms. Note, that this value of the decay rate can be
made even larger e.g. by tuning the scattering length $a_s$ via a Feshbach resonance, as $\Gamma \propto
a_s^2$, by increasing the density of the external gas or by increasing the lattice depth.

In summary we have shown that for a cold reservoir gas with sufficiently many atoms fast ground state cooling
of lattice atoms can be achieved with the dissipative coupling of the lattice to the reservoir. The necessary
experimental parameters have already been achieved in real experiments, and the cooling rates are tunable via
the scattering length and the density of the reservoir gas.

\section{Combined Process}\label{section:combined}

The combination of the cooling process with laser-assisted loading in the limit $\Omega \ll \omega,
\epsilon_F$ will give a final high-fidelity state in the lowest motional level. The primary role of the
dissipative element is to transfer atoms into a state in which they are not coupled back to the Fermi
reservoir, which is made possible because of the selective addressing of the $n=1$ motional levels in this
regime. Multiple occupation of a single site in the lowest motional state is forbidden due to Pauli-blocking,
and thus the lowest motional state is monotonically filled; the filling factor and hence the fidelity of the
state being prepared always improving in time. Again, patterns of atoms may be loaded in the lowest state by
using a superlattice to shift the energy of the $n=1$ motional level out of resonance with the Raman process
in particular sites, preventing atoms from being coupled from the Fermi reservoir into those sites. This
energy shift will also further suppress tunnelling of atoms from neighbouring sites.

If the laser-assisted loading and the cooling are carried out separately, each being performed after the
other in iterative steps, then from the analysis of sections \ref{section:coherent} and
\ref{section:incoherent} we see that an arbitrarily high fidelity final state can be obtained. This
\textit{pulsed} scheme gives us an upper bound on the timescale for loading a state of given fidelity, which
corresponds to the combination of the two individual timescales for laser-assisted loading and cooling.
Provided that the number of atoms in the reservoir is much larger than the number of lattice sites to be
filled ($N\gg M$), and the Markov approximation made in describing the cooling dynamics is valid, then there
will be no adverse effects arising from the loading and cooling processes sharing the same reservoir. Thus,
we can combine the two processes into a \textit{continuous} scheme, which in practice will proceed much
faster, as the continuous evacuation of the excited band due to cooling will also speed up the loading
process.

At the end of the loading process we must still ensure that the finite occupation of the excited motional
levels is properly removed. This can be achieved by detuning the resonant frequency for the Raman coupling
above the Fermi energy after the loading sweep, coupling the remaining atoms to empty states above the Fermi
sea, and then switching off the coupling adiabatically.

The dynamics of the pulsed process are already well understood from the analysis of sections
\ref{section:coherent} and \ref{section:incoherent}. To illustrate the dynamics of the combined continuous
process, we again perform numerical simulations, in which we compute the matrix elements of the reduced
system density operator. The dynamics of the total system including both the laser coupling and the
collisional interaction between the optical lattice and the Fermi reservoir are described by the full
Hamiltonian
\begin{align}
    H=H_a+H_b+H_{\rm RC}+H_{\rm int},\label{H0}
\end{align}
and in the Markov approximation with respect to the cooling process, the matrix elements of the system
density operator can now be calculated from the Master equation (\ref{MEalbet}) as shown in Appendix
\ref{TEcombined}. In order to obtain a closed set of differential equations which can be integrated
numerically, we use an approximation based on Wick's theorem to factorize fourth order correlation functions
into second order correlation functions (see appendix \ref{TEcombined}) \cite{Simulation}.

\begin{figure}[htp]
\includegraphics{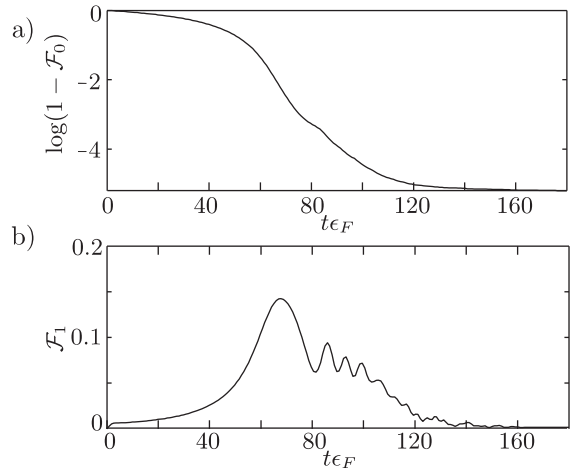}
\caption{The occupation of (a) the lowest ($n=0$) and (b) first excited ($n=1$) motional level for the
continuous combination of coherent loading in the slow regime and dissipative cooling. The resonant frequency
is swept from $\epsilon=0$ to $\epsilon=4\epsilon_F$ and the Raman coupling is switched off adiabatically.
Parameters used: $N=101$, $M=5$, $\Omega/2=0.45 \epsilon_F$, $n_{\rm 1D}\lambda/2=1.4$, $\omega=10
\epsilon_F$ and $\Gamma=0.1\epsilon_F$.}\label{Fig:combined}
\end{figure}

In Fig.~\ref{Fig:combined} we plot the time evolution of the occupation of the two motional levels in the
continuous regime as a function of time. In Fig.~\ref{Fig:combined}a we see that we indeed achieve a high
occupation of the lowest motional level from the combination of laser-assisted coupling to the excited
motional level in the regime $\Omega \ll \omega,\epsilon_F$ and cooling to the ground state. For the typical
values given in the figure caption, the loading time $T$ for a state with $\mathcal{F}_0\sim 1-10^{-4}$ is
again on the order of a few milliseconds. This required loading time can be further decreased by tuning
$\Gamma$ via the density of the external gas or the strength of the collisional interaction between atoms in
the lattice and atoms in the reservoir.

From Fig.~\ref{Fig:combined}b we see that as we fill the lower motional level, the filling in the upper level
is depleted, and as we continue to tune the lasers so that this level is coupled to states in the reservoir
above the Fermi energy $\varepsilon_F$, the remaining population in this level is removed.

As a final remark we note that such a procedure could, in principle, also be applied to bosons. However,
without Pauli blocking to prevent double-occupation of the ground motional level, we rely on the onsite
collisional shift $U$ to make the Raman coupling of an atom from the reservoir into an excited state
off-resonant if an atom already exists in the ground motional state. Second order processes occurring at a
rate $\sim \Omega\Gamma/U$ can still create double occupation, so we require $\Omega, \Gamma \ll U$, and the
advantage of true fault tolerance is not present as it is for fermions.

\section{Summary}

In conclusion, we have shown that the combination of laser-assisted loading of atoms into an excited motional
state and the cooling of atoms from this motional state to the ground level gives a fault-tolerant loading
scheme to produce high fidelity registers of fermions in an optical lattices with one atom per lattice site.
Application of a superlattice allows this to be extended to generalised patterns of atoms, and all of these
processes can be completed on timescales much faster than those on which atoms can be lost from the lattice.
The advantage of this scheme is that the dissipative transitions in the lattice, similar to optical pumping,
gives us a process in which the fidelity of the final state (in the lowest motional level) improves
monotonically in time.

\begin{acknowledgments}
The authors would like to thank Peter Rabl for helpful discussions. AG thanks the Clarendon Laboratory and DJ
thanks the Institute for Quantum Optics and Quantum Information of the Austrian Academy of Sciences for
hospitality during the development of this work. Work in Innsbruck is supported by the Austrian Science
Foundation, EU Networks, OLAQUI, and the Institute for Quantum Information. DJ is supported  by EPSRC through
the QIP IRC (www.qipirc.org) (GR/S82176/01) and the project EP/C51933/1.
\end{acknowledgments}

\appendix

\section{Derivation of the Heisenberg Equations for Coherent Loading}\label{Heisenberg}

Consider a system, which is described by a Hamiltonian quadratic in a set of operators $\vec{\mathcal{O}}=
(\mathcal{O}_1, \mathcal{O}_2, \dots,\mathcal{O}_d)$. Then the Heisenberg equations of motion can be written
as
\begin{align}\label{EoMHb}
    \dot{\vec{\mathcal{O}}}(t)=M\vec{\mathcal{O}}(t),
\end{align}
with a matrix $M$, and formal solution $\vec{\mathcal{O}} (t) = U\vec{\mathcal{O}}(0)$ with $U={\rm
exp}(Mt)$. By choosing the initial conditions $\mathcal{O}_j(0) =\delta_{j,\alpha}$ we can construct the full
time evolution matrix $U(t)$ by solving Eqs.~(\ref{EoMHb}), as
\begin{align}
    U_{i,\alpha}(t)\equiv \sum_j U_{i,j}(t) \mathcal{O}_j(0)=\mathcal{O}_i(t).
\end{align}
The time evolution of the second order correlation functions is then easily calculated as
\begin{align}
    \EV{\mathcal{O}_i^\dagger \mathcal{O}_j(t)}=\left\langle\sum_{\alpha,\beta} U_{i,j}^*(t)
    U_{j,\beta}(t) \mathcal{O}_\alpha^\dagger \mathcal{O}_\beta(0)\right\rangle.
\end{align}

\section{Derivation of the Master Equation}\label{ME}

In the interaction picture, and after making the Born-Markov approximation, the master equation for the
reduced density operator $\rho$ of a system which interacts with a heat bath via an interaction Hamiltonian
$H_{\rm int}$ can be written as (see e.g.~\cite{QN})
\begin{align}\label{ME0}
    \dot\rho(t)=-\int_0^t d\tau \rm{Tr_B}\Big\{\Big[ H_{\rm int}(t),\left[ H_{\rm
    int}(t-\tau),\rho(t)\otimes\rho_B\right]\Big]\Big\}.
\end{align}
Here, $\rho_B$ is the bath density operator, and $\rm{Tr_B}$ denotes the trace over the bath, which is
represented by the cold Fermi reservoir in our setup. The interaction between the Fermi reservoir and the
optical lattice system is given by the Hamiltonian (\ref{Hint}), and in the interaction picture with respect
to the internal dynamics in the lattice and in the Fermi reservoir,
\begin{align}
    H_{\rm int}(t) = \sum_{\bk,\bk'\atop\alpha,\bn,\bn'} \g \Df{\bk}{\bk'}{\alpha}{\bn}{\bn'}
    \rme^{-\rmi (\epsilon_\bk-\epsilon_{\bk'} + \omega (n-n'))t}.
\end{align}

As the number of atoms in the reservoir exceeds the number of lattice sites, $N\gg M$, and as in addition the
the bath has temperature $T\sim 0$, the reservoir will approximately remain in its ground state, i.e., the
filled Fermi sea throughout the cooling process, and the bath correlation functions are approximately given
by
\begin{align}
    \EV{b_{\bk_1}^\dagger b_{\bk_1'} b_{\bk_2}^\dagger b_{\bk_2'}} \approx \delta_{\bk_1,\bk_1'}
    \delta_{\bk_2,\bk_2'} +\delta_{\bk_1,\bk_2'}\delta_{\bk_1',\bk_2},
\end{align}
where $\EV{\,\cdot\,} = \rm{Tr_B} \{\,\cdot\,\rho_B\}$.

For $t$ much larger than the correlation time in the bath we can let the upper limit of the integral in
Eq.~(\ref{ME0}) go to $\infty$, and writing $\int_0^\infty \rme^{\rmi(\epsilon_\bk-\epsilon_{\bk'} + \omega
(n-n'))\tau} \rightarrow \delta(\epsilon_\bk-\epsilon_{\bk'} + \omega (n-n'))$, we find
\begin{align}\label{MEalbetappendix}
    \dot\rho=\sum_{\alpha,\beta,\bn}\frac{\Gamma_{\alpha,\beta,\bn}}{2} \left(2A_{\alpha,\bn} \rho
    A_{\beta,\bn}^\dagger -A_{\alpha,\bn}^\dagger A_{\beta,\bn} \rho- \rho A_{\alpha,\bn}^\dagger
    A_{\beta,\bn}  \right),
\end{align}
with the jump operator $A_{\alpha,\bn}= a_{\alpha,0}^\dagger a_{\alpha,\bn}$,
\begin{align}\label{Gammaappendix}
    \Gamma_{\alpha,\beta,\bn}=&2\pi \sum_{\bk, \bk' \atop k'>k} g_{\alpha,\bf{1},\bf{0}}^{\bk,\bk'}
    {g_{\beta,\bf{1},\bf{0}}^{\bk,\bk'}}^* \delta \left(\omega-\epsilon_\bk+ \epsilon_{\bk'} \right),
\end{align}
and where we note that $\sum_{\bk,\bk'} g_{\alpha,\bn,\bf{0}}^{\bk,\bk'} g_{\alpha,\bn',\bf{0}}^{\bk,\bk'} =
0$ for $\bn\not=\bn'$. The rate $\Gamma_{\alpha,\beta,\bn}$ rapidly decays with $|\alpha-\beta|$, and for
each of the three degenerate excited states $\bn\in \{(1,0,0),(0,1,0),(0,0,1)\}$, the slowest rate of this
decay is found in the direction of $\bn$. In the harmonic oscillator approximation we find (for the direction
with the slowest decay)
\begin{align}
    \Gamma_{\alpha,\beta}\sim\frac{g^2n_{3\rm D}m}{\pi a_0\sqrt{2}}\frac{2}{3e}
    F(\pi\sqrt{\frac{\omega}{\omega_R}}|\alpha-\beta|),
\end{align}
to first order in $\epsilon_F/\omega$, with the function
\begin{align}
    F(\xi)=3\frac{2\xi\cos\xi+(\xi^2-2)\sin\xi}{\xi^3}.
\end{align}
For large $\xi$ this result simplifies to the sinc function in Eq. (\ref{sincdecay}). For a deep optical
lattice where $\omega\gg\omega_R$, $F(\pi\sqrt{\frac{\omega}{\omega_R}}|\alpha-\beta|) \approx
\delta_{\alpha,\beta}$, and we end up with a standard quantum optical master equation (see e.g.~\cite{QN}),
describing the decay of an excited lattice atom from each of the three degenerate $n=1$ states to the $n=0$
level at a rate $\Gamma$.

\section{Equations of motion for Combined Dynamics}\label{TEcombined}

The time evolution of the expectation value of an arbitrary system operator $\hat{\mathcal{O}}$ can be
calculated from the master equation (\ref{MEalbet}) and Eq.~(\ref{Gamma}) as
\begin{align}\label{ME_op}
    \EV{\dot{\hat{\mathcal{O}}}} =&\rmi\EV{[H_{\rm sys},\hat{\mathcal{O}}]}+\nonumber\\
    &+\frac{\Gamma}{2}\sum_{\alpha,\bn} \left(
    2\EV{A_{\alpha,\bn}^\dagger\hat{\mathcal{O}} A_{\alpha,\bn}} - \EV{\{\hat{\mathcal{O}},A_{\alpha,\bn}^\dagger A_{\alpha,\bn} \}}
    \right),
\end{align}
where $H_{\rm sys}=H_a+H_b+H_{\rm RC}$ and $\Gamma\equiv\Gamma_{\alpha,\alpha,\bf{1}}$. We are interested in
the time evolution of the matrix elements of the single particle density matrix, which can be calculated from
Eq.~(\ref{ME_op}) as
\begin{widetext}
\begin{eqnarray}\label{TEMEfull}
    &\dt\EV{a_{\alpha,0}^\dagger a_{\beta,0}} = &\rmi\frac{\Omeff}{2}\sum_\bq \left(
    R_{\bq,\alpha,0} \EV{b_\bq^\dagger a_{\beta,0}}-R_{\bq,\beta,0}^* \EV{a_{\alpha,0}^\dagger b_\bq}\right)
    +\frac{\Gamma}{2}\left(2\EV{a_{\alpha,1}^\dagger a_{\alpha,1}}\delta_{\alpha,\beta}-
    \EV{a_{\alpha,1}^\dagger a_{\alpha,0}^\dagger a_{\beta,0}a_{\alpha,1}}-
    \EV{a_{\beta,1}^\dagger a_{\alpha,0}^\dagger a_{\beta,0}a_{\beta,1}} \right)\nonumber\\
    &\dt\EV{a_{\alpha,1}^\dagger a_{\beta,1}} = &\rmi\frac{\Omeff}{2}\sum_\bq \left(
    R_{\bq,\alpha,1} \EV{b_\bq^\dagger a_{\beta,1}}-R_{\bq,\beta,1}^* \EV{a_{\alpha,1}^\dagger b_\bq}\right)
    -\frac{\Gamma}{2}\left(2\EV{a_{\alpha,1}^\dagger a_{\beta,1}}-
    \EV{a_{\alpha,1}^\dagger a_{\beta,0}^\dagger a_{\beta,0}a_{\beta,1}}-
    \EV{a_{\alpha,1}^\dagger a_{\alpha,0}^\dagger a_{\alpha,0}a_{\beta,1}} \right)\nonumber\\
    &\dt\EV{a_{\alpha,1}^\dagger a_{\beta,0}} = &\rmi(\omega-\epsilon)\EV{a_{\alpha,1}^\dagger a_{\beta,0}}+
    \rmi\frac{\Omeff}{2}\sum_\bq \left( R_{\bq,\alpha,1} \EV{b_\bq^\dagger a_{\beta,0}} - R_{\bq,\beta,0}^*
    \EV{a_{\alpha,1}^\dagger b_\bq}\right) \nonumber\\
    &&+\frac{\Gamma}{2}\left(-\EV{a_{\alpha,1}^\dagger a_{\beta,0}}-
    \EV{a_{\alpha,1}^\dagger a_{\beta,1}^\dagger a_{\beta,1} a_{\beta,0}}-
    \EV{a_{\alpha,1}^\dagger a_{\alpha,0}^\dagger a_{\alpha,0}a_{\beta,0}} \right)\nonumber\\
    &\dt\EV{a_{\alpha,1}^\dagger b_\bk} = &\rmi(\epsilon-\epsilon_\bk)\EV{a_{\alpha,1}^\dagger b_\bk}+
    \rmi\frac{\Omeff}{2}\left( \sum_\bq R_{\bq,\alpha,1} \EV{b_\bq^\dagger b_\bk} -  \sum_\mu R_{\bk,\mu,1}
    \EV{a_{\alpha,1}^\dagger a_{\mu,1}}\right)- \frac{\Gamma}{2}\left(\EV{a_{\alpha,1}^\dagger b_\bk}+
    \EV{a_{\alpha,1}^\dagger a_{\alpha,0}^\dagger a_{\alpha,0} b_\bk} \right)\nonumber\\
    &\dt\EV{a_{\alpha,0}^\dagger b_\bk} = &\rmi(\epsilon-\epsilon_\bk-\omega)\EV{a_{\alpha,0}^\dagger b_\bk}+
    \rmi\frac{\Omeff}{2}\left( \sum_\bq R_{\bq,\alpha,0} \EV{b_\bq^\dagger b_\bk} -  \sum_\mu R_{\bk,\mu,0}
    \EV{a_{\alpha,0}^\dagger a_{\mu,0}}\right)- \frac{\Gamma}{2} \EV{a_{\alpha,1}^\dagger a_{\alpha,0}^\dagger
    b_\bk a_{\alpha,1} }\nonumber\\
    &\dt\EV{b_\bk^\dagger b_{\bk'}} = &\rmi(\epsilon_\bk-\epsilon_{\bk'}-\omega)\EV{b_\bk^\dagger b_{\bk'}}+
    \rmi\frac{\Omeff}{2}\sum_{\mu,n} \left(-R_{\bk,\mu,n} \EV{b_\bk^\dagger a_{\mu,n}} - R_{\bk,\mu,n}^*
    \EV{a_{\mu,n}^\dagger b_{\bk'}}\right).
\end{eqnarray}
\end{widetext}
A closed set of equations can be obtained from Eqs.~(\ref{TEMEfull}) by using Wick's theorem to factorize
fourth order correlation functions into products of second order correlation functions according to
\begin{align}
    \EV{\hat{c_1}\hat{c_2}\hat{c_4}\hat{c_4}}=\EV{\hat{c_1}\hat{c_2}}\EV{\hat{c_3} \hat{c_4}} -
    \EV{\hat{c_1}\hat{c_3}} \EV{\hat{c_2}\hat{c_4}}+ \EV{\hat{c_1}\hat{c_4}}
    \EV{\hat{c_2}\hat{c_3}},\nonumber
\end{align}
for fermionic operators $\hat{c_i}\in \{a_{\alpha,\bn}^\dagger, a_{\alpha,\bn},b_\bk^\dagger, b_\bk\}$ (see
e.g.~\cite{Castin_BEC}).

\end{document}